\documentclass[prl,twocolumn,superscriptaddress,showpacs,floatfix]{revtex4-1}
\usepackage[T1]{fontenc}
\usepackage{amsmath,amsfonts,upgreek}
\usepackage{graphicx}

\begin{document}
\title{Optical gap solitons and truncated nonlinear Bloch waves in temporal lattices}

\author{Christoph Bersch}%
\affiliation{Institute of Optics, Information and Photonics, University Erlangen-Nuremberg, Staudtstr. 7/B2, 91058 Erlangen, Germany}
\affiliation{Max Planck Institute for the Science of Light, Guenther-Scharowsky-Str. 1/B24, 91058 Erlangen, Germany}

\author{Georgy Onishchukov}%
\affiliation{Max Planck Institute for the Science of Light, Guenther-Scharowsky-Str. 1/B24, 91058 Erlangen, Germany}
\author{Ulf Peschel}%
\affiliation{Institute of Optics, Information and Photonics, University Erlangen-Nuremberg, Staudtstr. 7/B2, 91058 Erlangen, Germany}

\begin{abstract}
  We experimentally demonstrate the formation and stable propagation of various
  types of discrete temporal solitons in an optical fiber system. Pulses
  interacting with a time-periodic potential and defocusing nonlinearity are
  shown to form gap solitons and nonlinear truncated Bloch waves. Multi-pulse
  solitons with defects, as well as novel structures composed of a strong
  soliton riding on a weaker truncated nonlinear Bloch wave are shown to
  propagate over up to eleven coupling lengths. The nonlinear dynamics of all
  pulse structures is monitored over the full propagation distance which
  provides detailed insight into the soliton dynamics.
\end{abstract}


\maketitle

The complex interplay between nonlinearity and periodicity determines the
dynamics of many physical systems and leads to the formation of self-localized
excitations. These so-called discrete solitons \cite{Flach_Discrete2008,
  Campbell_Localizing2004} are subject of active research in many areas of
physics such as Bose--Einstein condensates \cite{Trombettoni_Discrete2001,
  Eiermann_Bright2004, Morsch_Dynamics2006} and nonlinear optics
\cite{Lederer_Discrete2008}. Especially nonlinear optical lattices provide a
prolific environment for exploring the peculiar effects of light propagation in
periodic systems \cite{Sukhorukov_Spatial2003, Christodoulides_Discretizing2003}
but have also been envisaged for networking and routing in all-optical circuits
\cite{Aceves_Discrete1996, Neshev_Controlled2004, Keil_All2011}.

Transferring these well-established concepts from space to equivalent systems in
time domain \cite{Treacy_Optical1969} gives access to completely new physical
phenomena related to the much richer spectral properties of temporal
systems. This may open up fundamentally new possibilities for engineering of
optical communication networks \cite{Peschel_Discreteness2008,
  Bersch_Spectral2011}. Exploiting the fast fiber nonlinearity paves the way for
studying many types of discrete phenomena. The sign of group velocity dispersion
allows to select between either focusing or defocusing nonlinearity
\cite{Peschel_Discreteness2008}.

Recent studies of spatial systems with defocusing nonlinearity confirmed
experimentally the existence of stable extended soliton clusters
\cite{Anker_Nonlinear2005, Alexander_Self2006, Wang_Truncated2009,
  Bennet_Observation2011} which had already been predicted earlier as
flat-top solitons \cite{Darmanyan_Discrete1999,
  Alexander_Soliton2006}. For a given set of parameters these truncated
nonlinear Bloch waves can occupy an arbitrary number of lattice sites
\cite{Wang_Truncated2009} and can take very complex forms
\cite{Alexander_Self2011}; their temporal analogs are archetypes of data
patterns in optical telecommunication.

In this Letter, we report on the first experimental observation of discrete
temporal solitons. The formation and stable propagation of fundamental gap
solitons and truncated nonlinear Bloch waves (TBW) mediated by a defocusing
nonlinearity is demonstrated in a temporal lattice using a recirculating
fiber-loop setup. Almost arbitrary bit patterns can be encoded and stabilized by
inserting internal defects into a TBW. The stable propagation of these new
structures is demonstrated over several coupling lengths. Finally, we study the
interaction of a single discrete soliton with a TBW and show experimentally the
robust propagation of this new collective state as well as its break-up
depending on the power levels of its components. A high-resolution all-optical
oscilloscope enables us to measure the soliton dynamics over the complete
propagation distance.

\begin{figure}
  \includegraphics{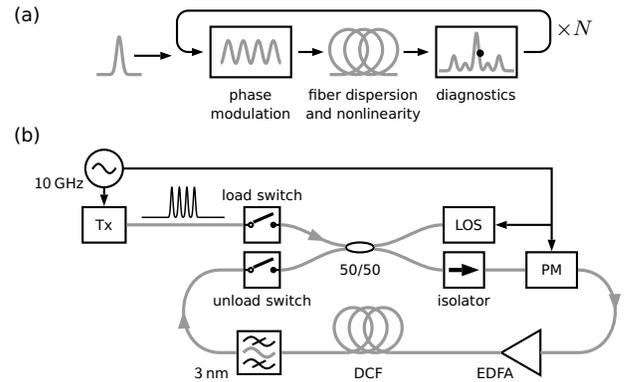}
  \caption{(a) Illustration of the experimental separation of phase
    modulation from fiber effects. (b) Block scheme of the experimental
    setup. A transmitter (Tx) generates patterns of 25\,ps pulses with
    10\,GHz pulse repetition rate at 1550\,nm. Acousto-optic load and
    unload switches control the loop operation. In each circulation a
    high-frequency periodic phase modulation (PM) is imposed on the
    pulses which are subsequently amplified with an Erbium-doped fiber
    amplifier (EDFA) before entering a dispersion-compensating fiber
    (DCF, $L = 1.4\,$km, $\gamma = 7\,\mathrm{(W\,km)}^{-1}$, $\beta_2 =
    120\,\mathrm{ps}^2/\mathrm{km}$, the total dispersion of all other
    components can be neglegted). An optical bandpass filter removes
    excess noise from the signal. The fiber loop circulation time is
    synchronized with the 10\,GHz microwave signal applied to the phase
    modulator. The pulse propagation is monitored with a high-resolution
    linear optical sampling setup (LOS).\label{fig:exp-setup}}
\end{figure}

We study pulse propagation in a recirculating fiber-loop setup which is
typically used to investigate optical long-haul transmission lines in the
lab. In our case, it consists of several kilometers of single-mode optical fiber
with normal group velocity dispersion, an Erbium-doped fiber amplifier to
compensate for the loop losses and a harmonically driven phase modulator (see
Fig.~\ref{fig:exp-setup}(b)).

The phase modulation provides a time-periodic potential which is applied
discretely once in each loop circulation \cite{Bersch_Spectral2011}, thereby
separating its action from fiber dispersion and nonlinearity, as it is
illustrated in Fig.~\ref{fig:exp-setup}(a). Here we deal only with phase
modulation much smaller than $2\uppi$ per round trip. Similar as for
guiding-center solitons \cite{Hasegawa_Guiding1991} and numerical split-step
algorithms \cite{Agrawal_Nonlinear1995}, a quasi-continuous model can be
applied.

The desired pattern consisting of 25\,ps pulses at 10\,GHz repetition rate is
injected via a 50\%~coupler into the fiber loop. After each round trip half of
the signal continues its propagation inside the loop, the remainder is coupled
out for monitoring. The measurements are performed with a linear optical
sampling setup \cite{Dorrer_Linear2003} which enables us to record the nonlinear
evolution of the signal power profile with high temporal resolution over the
complete propagation distance. Such a detailed experimental insight in nonlinear
pulse propagation in discrete systems is, at present, unattainable in equivalent
spatial arrangements \cite{Barsi_Imaging2009, Lederer_Discrete2008,
  Morsch_Dynamics2006}.

The propagation of picosecond pulses in our optical fiber system is well
described by a modified Nonlinear Schr\"odinger Equation
\cite{Wabnitz_Suppression1993, Agrawal_Nonlinear1995, Peschel_Discreteness2008}
\begin{equation}
  \label{eq:base-prop}
  \text{i}\partial_ZA - \frac{\beta_2}{2}\partial_T^2A + \gamma|A|^2A + V_0 \sin^2(\pi T/T_0) A = 0,
\end{equation}
where $T = t_\text{lab} - Z/v_g$ is time in the reference frame co-moving with
the pulse envelope $A$ at its group velocity $v_g$ and $Z$ is the propagation
coordinate. $V_0$ is the amplitude of an effective time-periodic potential with
period $T_0$ which depends on the phase modulation depth $\Phi_0$ and the length
of the loop $L_0$ as $V_0=\Phi_0/L_0$. $\beta_2$ is the group-velocity
dispersion (GVD) and $\gamma$ the Kerr nonlinearity of the fiber (see
Fig.~\ref{fig:exp-setup} for experimental parameters). Normalizing
Eq.~(\ref{eq:base-prop}) to characteristic scales gives us
\begin{equation}
  \label{eq:norm-prop}
  \text{i}\partial_zU - \sigma\partial_t^2U + |U|^2U + N_0\sin^2(t) U = 0\,,
\end{equation}
where $t = \pi T/T_0$, $z = Z/Z_0$ with $Z_0 = 2T_0^2/(\pi^2|\beta_2|)$, $U =
\sqrt{Z_0\gamma}A$, and the potential strength is scaled as $N_0 = Z_0 V_0$. The
coupling between the sites of the temporal lattice is facilitated by the GVD and
can be positive ($\sigma = +1$) for normal or negative ($\sigma = -1$) for
anomalous dispersion, in contrast to analogous spatial systems where diffraction
restricts the coupling to positive values. For normal GVD as studied here,
Eq.~(\ref{eq:norm-prop}) is equivalent to well-investigated spatial systems with
defocusing nonlinearity \cite{Lederer_Discrete2008, Wang_Truncated2009}.

\begin{figure}
  \includegraphics{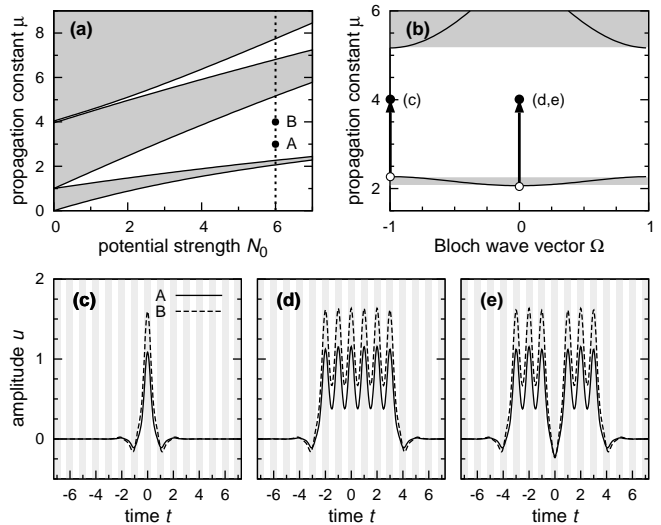}
  \caption{(a) Dependence of the linear transmission bands (shaded) on
    the potential depth $N_0$. (b) shows the bifurcation behavior of the
    numerical soliton solutions \cite{Yang_Universally2007a} in (c)--(e)
    with respect to the band structure of Bloch waves ($\Omega$:
    normalized Bloch vector). (c) shows a fundamental gap soliton, (d) a
    six-peak TBW, and (e) a six-peak TBW with defect. The soliton
    solutions are shown for a fixed potential strength $N_0 = 6$ at
    propagation constants $\upmu = 3.0$ (solid line) and $\upmu=4.0$
    (dashed line), which are also marked in the band structure (a) as A and
    B, respectively.}
  \label{fig:bs}
\end{figure}

Equation~(\ref{eq:norm-prop}) supports stationary solitary wave
solutions of the form $U(z,t) = u(t)\exp(\text{i}\upmu z)$. Among them
are clusters known as flat-top solitons or TBW
\cite{Wang_Truncated2009,Alexander_Self2006,
  Darmanyan_Discrete1999,Alexander_Soliton2006} which can be viewed as a
composition of fundamental gap solitons \cite{Zhang_Composition2009}.
Figure~\ref{fig:bs}(b) illustrates the bifurcation behavior of the
solitons with respect to the band structure of Bloch waves. TBW like the
one shown in Fig.~\ref{fig:bs}(d) do not bifurcate from the first band
like fundamental gap solitons, but when increasing the power the
defocusing nonlinearity shifts them out of the first band before they
localize in the first band gap \cite{Alexander_Self2006,
  Wang_Truncated2009, Bennet_Observation2011}. A TBW which features an
internal defect, as is displayed in Fig.~\ref{fig:bs}(e), belongs to an
own soliton family as it has a distinct topological structure. All these
soliton compositions can be excited experimentally as we will
demonstrate in the following. Solitons residing in other band gaps
\cite{Zhang_Composition2009} or for anomalous dispersion do also exists
\cite{Alexander_Soliton2006} and are also expected to be accessible
experimentally. All the created localized structures are completely
immobile and localize on individual lattice sites. This is different
from gap solitons observed in Bragg gratings, which cover hundreds of
unit cells and can even move across the lattice.

\begin{figure}
  \includegraphics{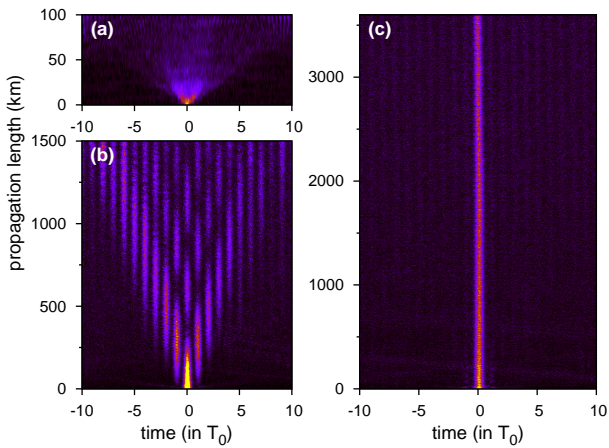}
  \caption{(Color online). Linear and nonlinear evolution of a single pulse. (a)
    shows the nonlinear pulse broadening without periodic potential and 30\,mW
    pulse peak power. (b) displays the discrete diffraction of the pulse in the
    temporal lattice for low power ($500\,\upmu$\kern-0.1em W), and (c) the
    discrete soliton with 30\,mW pulse peak power. In both cases the phase
    modulation depth is $0.5$~rad ($N_0 = 6$).}
  \label{fig:exp-1}
\end{figure}

Typical experimental results for linear and nonlinear evolution of
single pulses are illustrated in Fig.~\ref{fig:exp-1}. Without any phase
modulation pulses spread quickly, as can be seen in
Fig.~\ref{fig:exp-1}(a), a process which is even accelerated by the
nonlinearity of the fiber \cite{Agrawal_Nonlinear1995}.  As soon as the
phase modulation is switched on, fields become localized at the phase
minima and the spreading slows down considerably. This discrete temporal
diffraction \cite{Bersch_Spectral2011} is equivalent to its spatial
counterpart observed in waveguide arrays. Using the maximum pulse
spreading angle $\alpha_\text{max}$ in the linear case,
Fig.~\ref{fig:exp-1}(b), we estimated the coupling length
\cite{Eisenberg_Diffraction2000} to $L_\text{cpl} \approx
T_0\pi/\alpha_\text{max} \approx 500$\,km for a phase modulation depth
of $\Phi_0 = 0.5$~rad corresponding to a potential strength of $N_0 = 6$
in the continuous model (Eq.~\ref{eq:norm-prop}). This value for the
potential strength is used for all measurements presented in this
Letter. The input pulses for the linear propagation shown in
Fig.~\ref{fig:exp-1}(b) have a peak power of $500\,\upmu$\kern-0.1em W. 
All power values are given as peak powers and are averaged over the
fiber length of one loop round trip to take into account the fiber
losses.

When increasing the power the propagation constant $\upmu$ of the field enters
the first band gap and a fundamental gap soliton forms. The soliton is localized
deep inside the gap for already 30\,mW (see Fig.~\ref{fig:exp-1}(c)). Note, that
this soliton peak power is orders of magnitude smaller than for any other
optical system supporting discrete solitons based on a fast nonlinearity
\cite{Lederer_Discrete2008}. We would like to emphasize that the bright solitons
form in a regime with strong normal GVD, which even enhances the nonlinear pulse
spreading in the absence of a supporting periodic potential, as can be seen in
Fig.~\ref{fig:exp-1}(a).

\begin{figure}
  \includegraphics{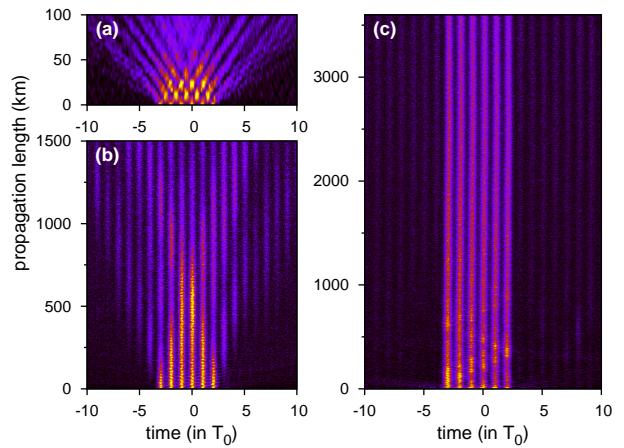}
  \caption{(Color online). Linear and nonlinear evolution of a six-peak pulse
    sequence. (a) shows the nonlinear broadening without temporal potential. The
    linear evolution with the lattice and the resulting discrete diffraction
    pattern is shown in (b). The formation and stable propagation of truncated
    nonlinear Bloch waves can be seen in (c). All parameters are as indicated in
    Fig.~\ref{fig:exp-1}.}
  \label{fig:exp-6}
\end{figure}

We could demonstrate stable soliton propagation over a distance of
3500\,km (2500 loop round trips) corresponding to seven coupling
lengths, as is demonstrated in Fig.~\ref{fig:exp-1}(c).  As each round
trip incorporates an amplification process, noise is added in form of
amplified spontaneous emission to the signal. Although this should also
result in Gordon--Haus timing jitter of the signal pulses
\cite{Gordon_Random1986, Kivshar_Optical2003a}, measurements like those
displayed in Fig.~\ref{fig:exp-1}(c) do not show noteworthy timing
fluctuations. This can be explained from two distinct perspectives: From
the viewpoint of discrete dynamics, fundamental gap solitons are
transversely immobile \cite{Aceves_Storage1994, Lederer_Discrete2008}
which manifests as timing stabilization in our setup. From a more
technical perspective, our experimental arrangement reminds of
synchronous modulation like it is used for retiming in all-optical
regenerators \cite{Zhu_102006}. Still, accumulation of amplified spontaneous
emission from optical amplifiers represents a major limitation for the
achievable propagation distance.

The optical transmitter allows us to generate and propagate arbitrary bit
patterns at 10\,GHz pulse repetition rate which is employed to study truncated
nonlinear Bloch waves. Figure~\ref{fig:exp-6} shows experimental results after
launching a sequence of six in-phase pulses into the fiber loop. Their linear
propagation inside the lattice results in spreading of the initial distribution
because of evanescent coupling. The maximum spreading angle imposed by the
periodic potential is clearly visible in Fig.~\ref{fig:exp-6}(b). The nonlinear
evolution in presence of the temporal lattice gives rise to stable TBW, as can
be seen in Fig.~\ref{fig:exp-6}(c). The experimental parameters are the same as
for the fundamental gap soliton in Fig.~\ref{fig:exp-1}.

\begin{figure}
\includegraphics{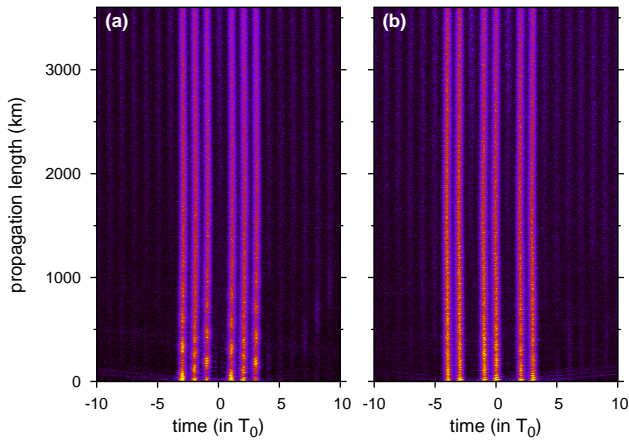}
\caption{(Color online). Nonlinear evolution of two pulse patterns with one (a)
  or two (b) single internal defects for $\hat{P} = 30\,$mW and $N_0 = 6$. The
  patterns form defect TBW which are stable for over 3500\,km.}
  \label{fig:exp-6defs}
\end{figure}

Moving towards more complex soliton states, we study the evolution of different
multi-pulse patterns featuring internal defects. When a single defect is
introduced into the six-peak pattern of Fig.~\ref{fig:exp-6}, the resulting
pulse sequence stays unchanged upon nonlinear propagation in the temporal
lattice (Fig.~\ref{fig:exp-6defs}(a)). This defect TBW belongs to an own soliton
family which is distinct from a mere combination of two three-peak solutions
(see e.g. \cite{Flach_Discrete2008}). Figure~\ref{fig:exp-6defs}(b) shows the
realization of another kind of defect TBW which features two single defects
separated by two lattice sites. These two patterns are representatives for
arbitrary bit patterns which can form solitons in the system.

\begin{figure}
\includegraphics{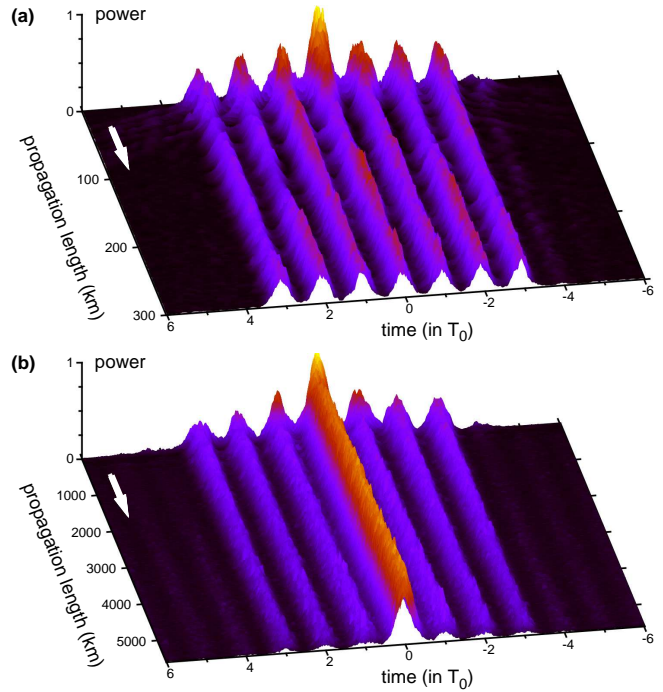}
\caption{(Color online). Measurement of nonlinear evolution of two pulse
  patterns consisting of seven peaks, with the central one having a peak power
  higher than the surrounding TBW. For a power ratio $r = 2.5$, the central
  pulse couples to the underlying TBW and decays (a). Its propagation is shown
  for only 300~km to visualize the break-up. An increased power ratio of $r=4$
  leads to an effective decoupling of the two solitons and the input pulse
  structure stays unchanged for at least 5600\,km (b).}
  \label{fig:exp-7plus1}
\end{figure}

From the perspective of our temporal approach, the existence of these distinct
TBW families is equivalent to nonlinear stabilization of arbitrary bit patterns
with on-off keying, thus suggesting potential applications in optical
communication. Contrary to common soliton transmission
\cite{Kivshar_Optical2003a}, the duty cycle of the pulses in a TBW is very high,
about 50\%. This value is close to that used in modern transmission systems with
return-to-zero modulation formats. The spectral efficiency of TBW is very high
because their spectrum is as narrow as that of single solitons.

Potentially, our system is not limited to binary on-off formats. Multi-pulse
patterns featuring a single peak with power higher than the surrounding TBW can
also be launched. The nonlinear propagation and in particular the robustness of
such multi-level structures is expected to depend critically on the power ratio.

Figure~\ref{fig:exp-7plus1}(a) displays the nonlinear evolution of such a novel
structure consisting initially of a strong central pulse having $r=2.5$ times
the peak power of the surrounding TBW. The strong peak distributes its energy
over the surrounding pulses and completely disappears after only a few loop
circulations. An isolated pulse of the same power of 50\,mW would immediately
form a gap soliton as can be deduced from Fig.~\ref{fig:exp-1}(c). For these
power levels the propagation constants of the fundamental gap soliton and the
TBW are too close, such that phase matching causes an efficient energy transfer
between them during propagation. It is worth noting that still all the power
remains confined to the initially excited seven lattice sites which somehow form
a kind of nonlinear background completely decoupled from the rest of the
lattice.

Increasing the power ratio leads to an efficient decoupling of the single pulse
from the background TBW. A typical measurement with $r=4$ (single pulse power of
60\,mW) is shown in Fig.~\ref{fig:exp-7plus1}(b) which clearly demonstrates that
the composition of TBW with a ``piggyback'' gap soliton maintains its initial
shape for at least 5600\,km, which corresponds to eleven coupling lengths. This
is even more surprising because the coexistence of these two solitary structures
with different propagation constants results in a non-stationary, but
nevertheless well-localized state. This is the first time, to our knowledge,
that stable propagation of such a structure has been observed.

In conclusion, we have demonstrated the formation of temporal solitary
structures in an effectively time-discretized optical fiber system. The
interplay of a fast nonlinearity and a time-periodic potential was employed to
observe temporal gap solitons as well as truncated nonlinear Bloch waves with
and without internal defects.  The pulse propagation at milliwatt peak powers
with defocusing nonlinearity was monitored with high temporal resolution over up
to eleven coupling lengths. Finally, we reported on the joint propagation of a
truncated nonlinear Bloch wave with a ``piggyback'' gap soliton. It was
demonstrated that this novel structure is robust for appropriate choice of
optical power. The attained symbiosis of discrete optics and fiber-based optical
communications not only sheds new light on long-known techniques like
synchronous modulation \cite{Zhu_102006}, active optical buffering
\cite{Jones_Asynchronous1998}, and ultra-long-haul optical data transmission
\cite{Nakazawa_101991}, but also indicates new possibilities for all-optical
signal processing.

\begin{acknowledgments}
  The authors acknowledge fruitful discussions with A. Regensburger and
  financial support by the Deutsche Forschungsgemeinschaft (Research Unit 532
  and Cluster of Excellence Engineering of Advanced Materials) and the
  German-Israeli Foundation.
\end{acknowledgments}

\end{document}